\documentclass[pdflatex,sn-nature]{sn-jnl}% Style for submissions to Nature Portfolio journals
\usepackage{graphicx}%
\usepackage{multirow}%
\usepackage{amsmath,amssymb,amsfonts}%
\usepackage{amsthm}%
\usepackage{mathrsfs}%
\usepackage[title]{appendix}%
\usepackage{xcolor}%
\usepackage{textcomp}%
\usepackage{manyfoot}%
\usepackage{booktabs}%
\usepackage{algorithm}%
\usepackage{algorithmicx}%
\usepackage{algpseudocode}%
\usepackage{listings}%
\usepackage{siunitx}%
\usepackage{float}
\theoremstyle{thmstyleone}%
\theoremstyle{thmstyletwo}%

\theoremstyle{thmstylethree}%

\begin{document}

\title[Transmission-Mode MSI for Microcystis and Microcystin]{Transmission-Mode Multispectral Imaging for Estimation of \textit{Microcystis} spp. Cell Concentration and ELISA-Equivalent Microcystin Concentration in Water}


% Author 1
\author[1,2]{\fnm{J.A.S.T.} \sur{Jayakody}}
\email{senith@eng.pdn.ac.lk}
\equalcont{These authors contributed equally to this work.}

% Author 2
\author[3,4]{\fnm{G.} \sur{Thilakarathne}}
\email{gayathrit@eie.ruh.ac.lk}
\equalcont{These authors contributed equally to this work.}

% Author 3 - Corresponding author
\author*[1,2]{\fnm{G.A.S.L.} \sur{Ranasinghe}}
\email{e19312@eng.pdn.ac.lk}
\equalcont{These authors contributed equally to this work.}

% Author 4
\author[1,2]{\fnm{M.C.L.} \sur{De Silva}}
\email{e19463@eng.pdn.ac.lk}

% Author 5
\author[1]{\fnm{K.A.D.S.A.} \sur{Perera}}
\email{e17245@eng.pdn.ac.lk}

% Author 6
\author[1,4]{\fnm{G.M.R.I.} \sur{Godaliyadda}}
\email{roshang@eng.pdn.ac.lk}

% Author 7
\author[1,4]{\fnm{H.M.V.R.} \sur{Herath}}
\email{vijitha@eng.pdn.ac.lk}

% Author 8
\author[1,4]{\fnm{M.P.B.} \sur{Ekanayake}}
\email{mpbe@eng.pdn.ac.lk}

% Author 9
\author[5]{\fnm{A.D.S.N.P.} \sur{Athukorala}}
\email{sarangiathu@sci.pdn.ac.lk}

% Author 10
\author[5]{\fnm{H.A.D.R.P.} \sur{Sirimanna}}
\email{ranmalees@sci.pdn.ac.lk}

% Affiliation 1
\affil*[1]{
\orgdiv{Department of Electrical and Electronic Engineering,
Faculty of Engineering},
\orgname{University of Peradeniya},
\orgaddress{
\city{Peradeniya},
\postcode{20400},
\country{Sri Lanka}
}
}

% Affiliation 2
\affil[2]{
\orgname{Farbe Technologies (Pvt) Ltd},
\orgaddress{
\city{Peradeniya},
\postcode{20400},
\country{Sri Lanka}
}
}

% Affiliation 3
\affil[3]{
\orgdiv{Department of Electrical and Information Engineering},
\orgname{University of Ruhuna},
\orgaddress{
\street{Hapugala},
\city{Galle},
\postcode{80000},
\country{Sri Lanka}
}
}

% Affiliation 4
\affil[4]{
\orgdiv{Multidisciplinary AI Research Centre},
\orgname{University of Peradeniya},
\orgaddress{
\city{Peradeniya},
\postcode{20400},
\country{Sri Lanka}
}
}

% Affiliation 5
\affil[5]{
\orgdiv{Department of Botany, Faculty of Science},
\orgname{University of Peradeniya},
\orgaddress{
\city{Peradeniya},
\postcode{20400},
\country{Sri Lanka}
}
}

\abstract{Harmful algal blooms dominated by \textit{Microcystis} spp. threaten freshwater systems through proliferation and microcystin (MC) production. Conventional measurements of cyanobacterial abundance and MC concentration commonly require laboratory processing, specialized instrumentation, or assay reagents. Because MC concentration cannot always be reliably inferred from cell abundance, complementary optical approaches for estimating both quantities are of interest. This study proposes and evaluates a transmission mode multispectral imaging (MSI) framework for estimating \textit{Microcystis} spp. cell concentration and ELISA-equivalent MC concentration from culture derived dilution samples. A custom built MSI system acquiring twelve spectral bands from near-ultraviolet to near-infrared wavelengths measured transmitted light spectral responses. Following principal component analysis based dimensionality reduction, spectral variation was quantified using an aggregated Wasserstein distance descriptor for regression based prediction. Cell concentration estimation yielded $R^2=0.9819$ for calibration and $R^2=0.9754$ on held-out intermediate dilution levels, while ELISA-equivalent MC estimation yielded $R^2=0.9788$ for calibration and $R^2=0.9786$ on held-out intermediate dilution levels. These results demonstrate the feasibility of transmission-mode MSI as a dual output optical calibration framework for estimation of \textit{Microcystis} spp. cell concentration and ELISA-equivalent MC concentration in water.}


\maketitle

\section{Introduction}\label{}
Harmful algal blooms (HABs), particularly those driven by cyanobacteria, represent a growing threat to freshwater quality, aquatic ecosystems, and public health. Cyanobacterial blooms can degrade drinking water sources, impair recreational waters, disrupt aquatic food webs, and cause illness in humans and animals through the production of cyanotoxins~\cite{Villalobos2025}. In 2022, cyanobacterial HABs reported through the One Health Harmful Algal Bloom System in the United States were associated with 95 human illnesses and more than 102,000 animal illnesses across fifteen states, illustrating the scale of their ecological and public health impact~\cite{CDC2022OHHABS}. As the frequency, duration, and spatial extent of HABs continue to increase globally~\cite{Feng2024}, reliable and timely monitoring approaches are increasingly required to support early detection, risk assessment, regulatory decision making, and long term freshwater management.
Among bloom forming cyanobacteria, \textit{Microcystis} spp. are of particular concern because of their widespread occurrence and ability to produce microcystins (MCs). MC-LR is one of the most extensively studied MC congeners~\cite{MCLR}, and blooms of MC producing \textit{Microcystis} spp. therefore represent an important consideration for freshwater quality monitoring. The World Health Organization (WHO) recommends provisional guideline values for total MCs of 1~\si{\micro\gram\per\liter} for lifetime drinking water exposure, 12~\si{\micro\gram\per\liter} for short term drinking water exposure, and 24~\si{\micro\gram\per\liter} for recreational exposure~\cite{WHO}. Accordingly, regulatory guidance emphasizes monitoring both cyanobacterial abundance and toxin concentrations~\cite{EPA2019,Sanseverino2023JRC}. This dual monitoring requirement reflects the fact that cyanobacterial abundance provides information on bloom magnitude, whereas toxin measurements provide complementary information on the associated chemical hazard. Therefore, analytical approaches that can support rapid, sensitive, and practical assessment of both bloom abundance and associated cyanotoxin concentrations are valuable for water quality assessment and management ~\cite{Rome2021}.

A key challenge in \textit{Microcystis} spp. bloom assessment is that cell concentration and MC concentration are related but distinct monitoring targets. Importantly, cyanobacterial cell concentration does not necessarily correlate with MC concentration, indicating that cell abundance alone may not reliably represent the associated toxin concentration~\cite{nocorrelation, Baek2026Microcystis}. Cell concentration describes bloom magnitude, whereas MC concentration provides a more direct indicator of toxin related risk. MC cell quota and congener composition can vary with genotype composition and colony structure, while environmental factors such as nutrient availability, temperature, physiological state, light exposure, and competition with co-occurring phytoplankton can further alter MC production~\cite{heptapeptide,Merel2013,nocorrelation,Wang2013WaterResearch,Cui2024WaterResearch}. Consequently, cell counts or pigment based indicators alone may provide an incomplete basis for toxin risk assessment. This distinction highlights the importance of assessing cyanobacterial abundance and associated MC concentrations as separate monitoring targets, rather than relying on bloom biomass alone as an indicator of toxin concentration.

Conventional approaches for \textit{Microcystis} spp. cell quantification include microscopy~\cite{WHO2021,HotzelCroome1999}, hemocytometer counting~\cite{Thunyaporn2021}, flow cytometry~\cite{Zhang2023}, and polymerase chain reaction (PCR) based techniques~\cite{Kataoka2013}, and remain widely used. These methods provide valuable information but differ in analytical specificity, throughput, instrumentation requirements, and suitability for field or laboratory applications~\cite{Jin2018,Zhang2023}. In situ fluorescence sensors provide a faster alternative for bloom detection, although their continuous submerged deployment can require regular maintenance to address optical window fouling and associated measurement drift~\cite{Bertone2018}. Thus, method selection depends on the monitoring objective and deployment context.

MC quantification is commonly performed using analytical and biochemical methods such as enzyme-linked immunosorbent assay (ELISA), protein phosphatase inhibition assay (PPIA), high-performance liquid chromatography (HPLC), and liquid chromatography-tandem mass spectrometry (LC-MS/MS)~\cite{method1,Sangolkar2006WaterResearch}. Their performance and practical requirements vary with the analytical method, sample matrix, target concentration, and monitoring context~\cite{method2}. ELISA is particularly useful because it provides sensitive quantification of total or equivalent MC concentrations, although antibody based responses may vary across MC congeners~\cite{method1,method2}. Rapid immunoassays can support operational screening, although response range, cross reactivity, matrix effects, and assay variability may affect their performance near decision thresholds~\cite{Humpage2012WaterResearch}.  Antibody based electrochemical biosensors have also been demonstrated for selective MC-LR detection~\cite{AdjeiNimoh2024}. Emerging biosensors offer additional opportunities for rapid toxin detection, but their reliance on immobilized biorecognition elements and specially fabricated sensing surfaces can present challenges related to storage stability, sensor regeneration, fabrication reproducibility, and scalable production \cite{Dai2025}. These complementary capabilities motivate the development of optical and data driven approaches that can support efficient estimation of cyanobacterial abundance and associated MCs. 

Optical sensing has increasingly been explored for cyanobacterial monitoring based on spectral responses. Optical approaches can be broadly classified as analyte specific sensing, in which a selective recognition mechanism or engineered sensing substrate produces an MC responsive signal, and indirect sensing based on non-specific optical proxies. Recent analyte specific approaches include MC-LR responsive near-infrared (NIR) fluorescent probes and surface enhanced Raman scattering (SERS) substrates for quantitative MC-LR detection~\cite{Chuan2025WaterResearch,Fan2026JECE}. In contrast, satellite and UAV based multispectral imaging (MSI) and hyperspectral imaging (HSI) primarily support bloom mapping and pigment retrieval~\cite{FANG2025123176,msi,Mishra2019Bloom}.  Laboratory hyperspectral studies have also characterized microalgal pigment signatures using absorbance or reflectance measurements~\cite{Naik2025Phytoplankton,TothNyari2026}. However, toxin related information is often inferred indirectly from pigment based proxies such as chlorophyll-\textit{a} or phycocyanin~\cite{Hunter2010Hyperspectral,Li2013ThreeBand,Pyo2019SAEDNN,msidis}. At the sample level, MSI similarly responds primarily to biomass, pigments, suspended particles, and scattering characteristics rather than directly detecting MCs. Consequently, MC estimation from MSI requires empirical calibration against independently measured MC concentrations, with its reliability depending on the robustness and transferability of the calibration relationship~\cite{Sun2024WaterResearch}.

Despite these advances, the potential of controlled, sample level MSI for estimating both cyanobacterial abundance and MC concentration remains insufficiently explored. In particular, it is necessary to investigate whether wavelength resolved transmission responses contain sufficient information to estimate cell concentration and ELISA-equivalent MC concentration, while preserving the distinction between these two monitoring targets. Accordingly, this study develops and evaluates a transmission mode MSI framework for estimating \textit{Microcystis} spp. cell concentration and ELISA-equivalent MC concentration in water samples. The framework combines twelve band MSI acquisition, spectral preprocessing, principal component analysis (PCA) based dimensionality reduction, and an aggregated Wasserstein distance descriptor. Two separate calibration relationships are established using independently measured reference cell concentrations and ELISA-equivalent MC concentrations. The corresponding regression models are evaluated on held-out intermediate dilution levels excluded from model fitting.

\section{Results}
\label{sec:results}

\subsection{Spectral Response of \textit{Microcystis} spp. Samples}
\label{subsec:spectral_response}

\begin{figure*}
    \centering
    \includegraphics[width=\textwidth]{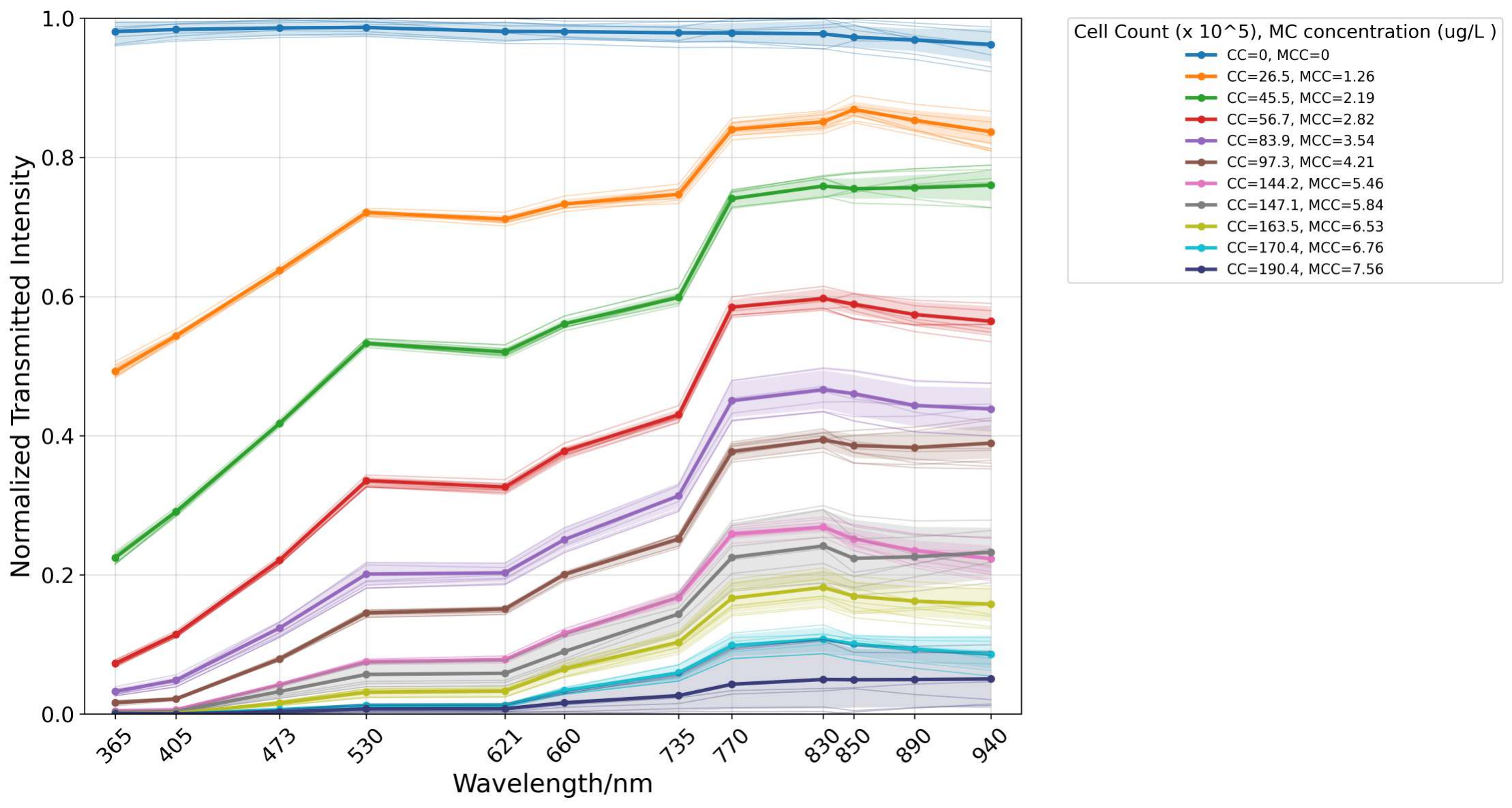}
    \caption{Normalized spectral signatures across the calibration concentration levels after preprocessing and dataset preparation.}
    \label{fig:spectral_sig}
\end{figure*}

Fig.~\ref{fig:spectral_sig} shows the normalized transmitted intensity profiles obtained across the twelve selected wavelength bands for the eleven calibration concentration levels. A systematic separation among the prepared concentration levels is observed across much of the measured spectral range, with normalized transmitted intensity generally decreasing as \textit{Microcystis} spp. cell concentration increases.
In the visible region, pronounced concentration dependent differences are observed across the selected bands between approximately 473 and 660~nm.
Concentration dependent separation is also evident in the NIR region from 735 to 940~nm.

\subsection{\textit{Microcystis} spp. Cell Concentration Estimation}
\label{subsec:cell_concentration_results}

The relationship between the aggregated Wasserstein distance descriptor and \textit{Microcystis} spp. cell concentration was modeled using the linear regression equation
\begin{equation}
\hat{C}_{\mathrm{cell}}
=
-0.1322
+
110.8299\mathcal{W},
\end{equation}
where $\mathcal{W}$ is the aggregated Wasserstein distance descriptor and $\hat{C}_{\mathrm{cell}}$ is the predicted cell concentration in units of $\times 10^5$ cells mL$^{-1}$.

\begin{figure}
    \centering
    \includegraphics[width=1\linewidth]{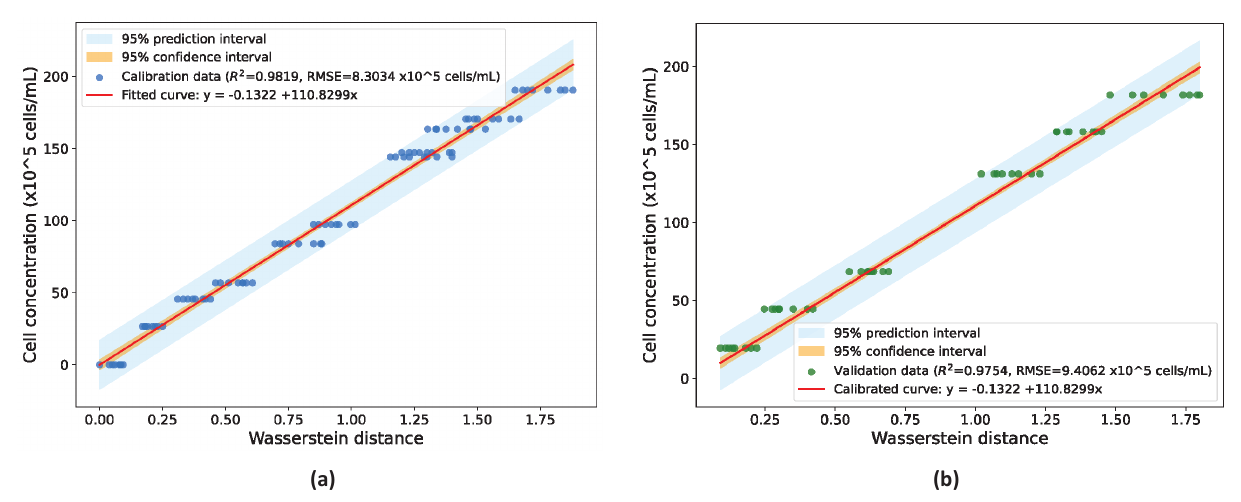}
    \caption{Calibration and held-out evaluation of \textit{Microcystis} spp. cell concentration estimation using the aggregated Wasserstein distance descriptor. (a) Calibration data comprising 88 MSI imaging replicates from eleven concentration levels. (b) Held-out validation data comprising 48 MSI imaging replicates from six intermediate concentration levels excluded from calibration.}
    \label{fig:cell_concentration}
\end{figure}

For the calibration set, the model achieved coefficient of determination, $R^2~=~0.9819$, with a root mean square error (RMSE) of $8.3034 \times 10^5$ cells mL$^{-1}$ and a mean absolute error (MAE) of $6.8740 \times 10^5$ cells mL$^{-1}$ (Fig.~\ref{fig:cell_concentration}(a)). When the same fitted model was applied without refitting to the held-out validation set, which contained intermediate concentration levels excluded from model fitting, performance remained comparable, with $R^2 = 0.9754$, an RMSE of $9.4062 \times 10^5$ cells mL$^{-1}$, and an MAE of $7.6229 \times 10^5$ cells mL$^{-1}$ (Fig.~\ref{fig:cell_concentration}(b)). The fitted relationship and uncertainty bands in both figures were estimated from the calibration data, and the held-out samples were evaluated without refitting.

The sample level uncertainty and error statistics for the cell concentration model are summarized in Table~\ref{tab:uncertainty_summary}. For the held-out validation set, the model showed a negative bias of $-3.5180 \times 10^5$ cells mL$^{-1}$, indicating average underprediction. The 95\% prediction interval coverage was $0.9167$, corresponding to 44 of the 48 held-out MSI sample level predictions having their assigned reference values within the calibration derived prediction intervals. These 48 MSI descriptors corresponded to six distinct held-out concentration levels.

\begin{table*}[htbp]
\centering
\caption{Sample level prediction performance and uncertainty summary for the proposed MSI Wasserstein calibration models. RMSE, MAE, bias, and residual standard deviation are reported in the corresponding target units. The 95\% prediction interval coverage indicates the proportion of MSI sample level predictions whose assigned reference values fell within the calibration derived prediction intervals.}
\label{tab:uncertainty_summary}
\resizebox{\textwidth}{!}{%
\begin{tabular}{llcccccc}
\hline
Target & Split & $R^2$ & RMSE & MAE & Bias & Residual SD & 95\% PI coverage \\
\hline
Cell concentration & Calibration & 0.9819 & 8.3034 & 6.8740 & 0.0000 & 8.3510 & 0.9773 \\
Cell concentration & Held-out validation & 0.9754 & 9.4062 & 7.6229 & -3.5180 & 8.8158 & 0.9167 \\
ELISA-equivalent MC concentration & Calibration & 0.9788 & 0.3408 & 0.2856 & 0.0000 & 0.3428 & 0.9773 \\
ELISA-equivalent MC concentration & Held-out validation & 0.9786 & 0.3362 & 0.2719 & -0.0739 & 0.3315 & 0.9792 \\
\hline
\end{tabular}%
}
\begin{flushleft}
\footnotesize{Note: Cell concentration errors are reported in $\times 10^5$ cells mL$^{-1}$, while ELISA-equivalent MC concentration errors are reported in $\mu$g L$^{-1}$. Calibration statistics are based on 88 MSI sample level descriptors associated with 11 concentration levels, while held-out statistics are based on 48 MSI sample level descriptors associated with six intermediate concentration levels.}
\end{flushleft}
\end{table*}

\subsection{ELISA-Equivalent Microcystin Concentration Estimation}
\label{subsec:mc_concentration_results}

ELISA-equivalent MC concentration was estimated from the aggregated Wasserstein distance descriptor using the fitted calibration relationship
\begin{equation}
\hat{C}_{\mathrm{MC}}
=
0.1849
+
4.3165\mathcal{W},
\end{equation}
where $\hat{C}_{\mathrm{MC}}$ denotes the predicted ELISA-equivalent MC concentration in $\mu$g L$^{-1}$ and $\mathcal{W}$ is the aggregated Wasserstein distance descriptor.

\begin{figure}
    \centering
    \includegraphics[width=1\linewidth]{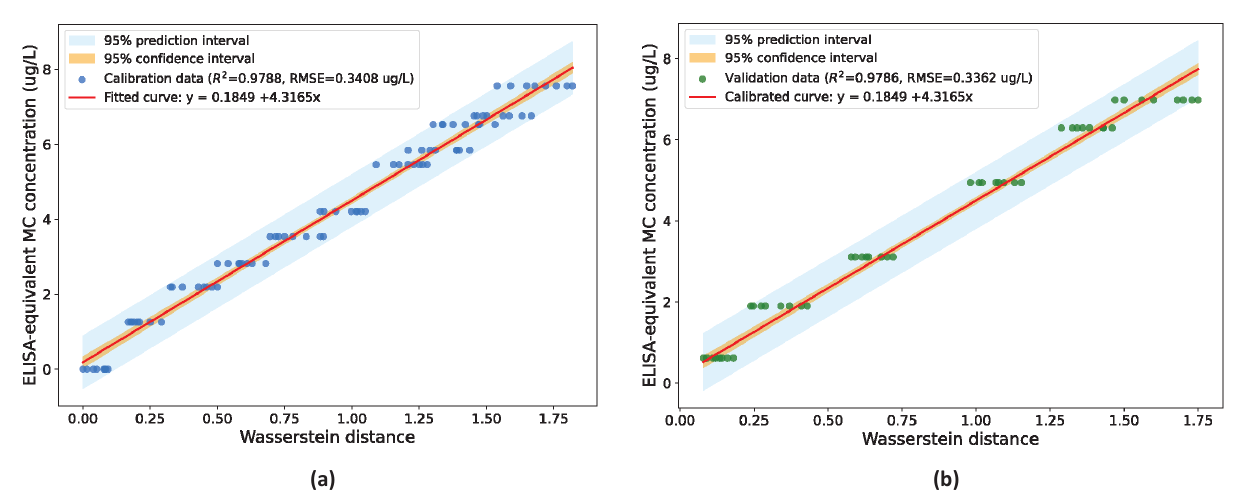}
    \caption{ELISA-equivalent MC concentration estimation using the aggregated Wasserstein distance descriptor. (a) Calibration data comprising 88 MSI imaging replicates from eleven concentration levels. (b) Held-out validation data comprising 48 MSI imaging replicates from six intermediate concentration levels excluded from calibration.}
    \label{fig:mc_concentration}
\end{figure}

The calibration relationship is presented in Fig.~\ref{fig:mc_concentration}(a), with $R^2 = 0.9788$, an RMSE of 0.3408~$\mu$g L$^{-1}$, and an MAE of 0.2856~$\mu$g L$^{-1}$. Application of this relationship to the held-out intermediate dilution levels yielded $R^2 = 0.9786$, while the RMSE and MAE were 0.3362 and 0.2719~$\mu$g L$^{-1}$, respectively (Fig.~\ref{fig:mc_concentration}(b)). Both the fitted relationship and its uncertainty bands were derived exclusively from the calibration data and applied unchanged during held-out evaluation.

Table~\ref{tab:uncertainty_summary} summarizes the associated error and uncertainty statistics. A held-out bias of $-0.0739$~$\mu$g L$^{-1}$ indicated slight average under prediction. The assigned reference concentrations fell within the calibration derived 95\% prediction intervals for 47 of the 48 held-out MSI samples, giving a coverage of $0.9792$. These 48 MSI descriptors corresponded to six distinct held-out ELISA-equivalent MC reference concentrations, with one level specific ELISA measurement assigned to the eight MSI imaging replicates at each concentration level.

\section{Discussion}
The present findings demonstrate the feasibility of using transmission mode MSI to estimate both \textit{Microcystis} spp. cell concentration and ELISA-equivalent MC concentration within a controlled culture derived dilution series. This dual target formulation avoids explicitly using one target as a proxy for the other and provides two complementary outputs from a single optical acquisition: an estimate of cyanobacterial cell concentration and an estimate of ELISA-equivalent MC concentration. In this study, both calibration relationships showed comparable performance on held-out intermediate concentration levels.

The decrease in normalized transmitted intensity with increasing cell concentration is consistent with increasing optical attenuation within the cyanobacterial suspension, arising from the combined effects of pigment related absorption and light scattering~\cite{Hunter2010Hyperspectral,Magri2023HSITransmittance}. The selected visible wavelengths overlap spectral regions influenced by chlorophyll-\textit{a} and the cyanobacterial accessory pigment phycocyanin~\cite{VISnNIR,Hunter2010Hyperspectral,Magri2023HSITransmittance,Li2013ThreeBand}. The observed decrease in transmitted intensity with increasing concentration is therefore consistent with greater optical attenuation as cyanobacterial biomass increases. However, because transmission mode measurements incorporate both absorption and scattering effects, the individual contributions of these mechanisms cannot be separated from the present MSI measurements alone. In the NIR wavelength range, the prominent visible region pigment absorption features are reduced, while scattering associated with suspended cells and colonies can contribute substantially to the measured optical response~\cite{Magri2023HSITransmittance}. The NIR bands therefore provide complementary information on concentration dependent optical changes in the suspension beyond those observed in the visible region. Overall, the multispectral response represents an integrated optical signature resulting from the combined wavelength dependent absorption and scattering characteristics of the samples.

The observed relationships are consistent with the spectral behavior described in Section~\ref{subsec:spectral_response}. Increasing cell concentration and ELISA-equivalent MC concentration were associated with progressively greater changes in the transmission mode spectral response relative to the non contaminated M8 medium reference as described in Section~\ref{subsec:experimental_sample_preparation}, reflecting the combined influence of wavelength dependent absorption and scattering within the cyanobacterial suspension. The aggregated Wasserstein distance descriptor summarizes these distributional changes across the four retained principal components into a single measure of spectral departure from the fixed M8 medium reference. The close agreement between the calibration and held-out results for both targets indicates that the fitted relationships remained stable when applied to intermediate concentration levels not used during model fitting. 

Previous optical approaches for cyanobacterial abundance estimation have generally followed three main routes: fluorescence imaging or microscopy based enumeration, hyperspectral transmission or reflectance analysis, and remote sensing based pigment retrieval. Fluorescence imaging approaches can provide accurate cyanobacterial cell quantification, but they generally require fluorescence instrumentation, cell level segmentation, or microscopy based workflows~\cite{Jin2018}. Hyperspectral transmission and reflectance studies provide substantially denser spectral information over broad wavelength ranges, resulting in higher dimensional datasets and more extensive spectral processing requirements~\cite{Hunter2010Hyperspectral,Pyo2019SAEDNN,Magri2023HSITransmittance}. Airborne and satellite based multispectral or hyperspectral approaches support large scale bloom mapping, but commonly retrieve pigment related quantities such as phycocyanin or chlorophyll-\textit{a} rather than directly enumerating cells or quantifying MC concentration~\cite{Hunter2010Hyperspectral,Li2013ThreeBand,Pyo2019SAEDNN,Sun2024WaterResearch}.

For MC quantification, analytical methods such as ELISA, HPLC, and LC-MS/MS remain established reference approaches~\cite{method1,method2}. Conventional direct spectrophotometric characterization of MCs commonly relies on UV region absorption at wavelengths substantially shorter than the 365-940~nm range used in the present MSI system~\cite{Sangolkar2006WaterResearch}. Analyte specific optical approaches using MC responsive fluorescent probes and engineered SERS sensing substrates have also demonstrated direct MC-LR detection at longer wavelengths~\cite{Chuan2025WaterResearch,Fan2026JECE}. Cell concentration and MC concentration nevertheless represent distinct monitoring quantities: the former characterizes bloom abundance, whereas the latter is more directly associated with toxin exposure. MC cell quota and congener composition can vary with colony size and genotype composition, while environmental factors can influence MC production~ \cite{Wang2013WaterResearch,Cui2024WaterResearch,Baek2026Microcystis}. Together with broader evidence on MC biology and bloom variability~\cite{heptapeptide,Merel2013,nocorrelation}, these findings indicate that cyanobacterial abundance or pigment related measurements alone may not reliably represent MC concentration.

In this context, the proposed framework demonstrates quantitative \textit{Microcystis} spp. cell concentration estimation and separate calibration of ELISA-equivalent MC concentration using twelve selected spectral bands in a transmission mode MSI configuration. Although previous studies differ in datasets, target variables, sensing geometries, and validation settings, the present results show that a reduced band MSI representation retains sufficient optical information for quantitative estimation within the tested dilution series. Some previous approaches use fewer bands, such as three band reflectance models for phycocyanin retrieval~\cite{Li2013ThreeBand}; however, such models generally target pigment concentration or bloom related indices rather than \textit{Microcystis} spp. cell concentration. The present framework provides both concentration estimates from the same optical acquisition through separately fitted calibration relationships. Because the measurement configuration uses no MC selective probe, recognition element, or engineered sensing substrate, the relationship with ELISA-equivalent MC concentration should be interpreted as an indirect empirical association within the controlled culture derived dilution series, rather than as direct spectroscopic detection of MC. The results therefore demonstrate the feasibility of using the MSI derived descriptor for ELISA-equivalent MC estimation.

More broadly, the results highlight the contribution of the complete analysis pipeline, including controlled illumination, selected wavelength acquisition, consistent imaging conditions, preprocessing, feature transformation, PCA based dimensionality reduction, aggregated Wasserstein distance descriptor construction, and regression modeling, to quantitative estimation from transmission mode MSI data. The compact LED based spectrophotometric measurements combined with data driven calibration can support rapid indirect estimation of water quality parameters~\cite{CarreresPrieto2023JECE}. Reference calibrated data driven soft sensors have also been investigated for estimating water quality indicators from indirect measurements~\cite{Mozo2022SoftSensor}. Although the target variables and sensing configuration differ from those of the present study, this illustrates the broader applicability of reference calibrated optical measurements for water quality characterization. The proposed methodology therefore provides an end-to-end framework for joint optical assessment of \textit{Microcystis} spp. abundance and ELISA-equivalent MC concentration. The held-out evaluation specifically assesses interpolation to intermediate concentration levels excluded from model fitting, rather than transfer across independently collected cultures or environmental matrices.

Further work should evaluate the framework using independently collected cultures, natural bloom samples, mixed cyanobacterial communities, and broader environmental and optical variability. Comparison with congener specific analytical measurements such as LC-MS/MS would also help establish the robustness and transferability of the approach for field oriented monitoring.

\section{Materials and Methods}
\subsection{Culture Isolation, Cultivation, and Growth Monitoring}
\label{subsec:culture_isolation}
A freshwater bloom sample containing \textit{Microcystis} spp. was provided by the Department of Botany, Faculty of Science, University of Peradeniya, Sri Lanka and subsequently processed in the laboratory for culture enrichment and propagation. Morphological identification was performed using an optical microscope (Olympus CX31, Japan) at $400\times$ magnification. The cyanobacterial material was identified as morphologically consistent with the genus \textit{Microcystis} spp.~\cite{WHO2021}. To obtain an isolated culture, the enriched material was subjected to serial dilution followed by streak plating on Bold's Basal Medium (BBM) agar (1.5\%)~\cite{Fernandez2021,UTEX2019BoldBasalMedium,Pretorius1981}. All media and glassware were sterilized by autoclaving at $121~^\circ$C for 45~min prior to use. The plates were incubated for 10-14 days at $25\pm2~^\circ$C under 50~$\mu$mol photons m$^{-2}$ s$^{-1}$ illumination with a 12:12~h light:dark photoperiod until discrete colonies formed~\cite{Shahmohamadloo2019}.

A well isolated colony was aseptically transferred from BBM agar into sterile liquid M8 medium~\cite{Mandalam1998} to prepare the inoculum. The cell concentration of the resulting suspension was determined by hemocytometer enumeration and adjusted to $1\times10^{5}$ cells mL$^{-1}$ before inoculation into a sterilized 8~L culture vessel containing M8 medium. The culture medium pH was adjusted to 8.0 prior to sterilization. The culture vessel was incubated under the same temperature, illumination, and photoperiod conditions. Continuous aeration at approximately 0.5~L~min$^{-1}$ was supplied using an aquarium pump to maintain suspension homogeneity and minimize cell aggregation.

Culture development was monitored for 30 days to characterize temporal changes in cell concentration. Aliquots of 10~mL were collected at selected time points for cell enumeration using a hemocytometer. The resulting growth profile was used to identify stationary phase~\cite{Chen2017,Huang2009} for preparation of the controlled concentration series, as shown in Supplementary Fig.~\ref{fig:growth_curve}.

\subsection{Experimental Sample Preparation}
\label{subsec:experimental_sample_preparation}
A controlled concentration series was prepared using stationary phase \textit{Microcystis} spp. culture grown in M8 medium to establish a calibration dataset and a held-out validation dataset comprising intermediate dilution levels. The stock culture was diluted with sterile M8 medium to obtain eleven calibration levels ranging from 0\% to 100\% (v/v) at 10\% intervals. The 0\% level, consisting of sterile M8 medium without added culture suspension, served as the medium matched reference control.

For each calibration level, a single bulk sample was prepared by combining the required volume of stationary phase stock culture with sterile M8 medium to obtain the designated dilution level. Each bulk preparation was thoroughly homogenized, and aliquots were subsequently collected for reference cell enumeration and ELISA-equivalent MC quantification. The aliquots designated for MC quantification were stored at $-20~^{\circ}$C prior to ELISA analysis. The cell concentration and ELISA-equivalent MC concentration determined from the respective reference aliquots were assigned as the level-specific reference values for the corresponding bulk preparation.

Following collection of the reference aliquots, the remaining homogenized bulk sample was apportioned into eight imaging replicates. Each imaging replicate was maintained separately and homogenized immediately prior to image acquisition. A single twelve band multispectral image stack was acquired from each imaging replicate using the same procedure. Thus, all eight imaging replicates at each concentration level originated from the same homogenized bulk preparation and shared the corresponding level-specific reference values.

A held-out evaluation set was established using six intermediate concentration levels of 5, 15, 35, 55, 75, and 95\% (v/v), which were not included in the calibration dataset. For each held-out validation concentration, bulk sample preparation, homogenization, reference aliquot collection, preparation of eight imaging replicates,  and MSI acquisition were performed following the procedures described above for the calibration samples. Reference cell concentrations were determined using the same hemocytometer based cell enumeration procedure (Section~\ref{subsec:cellCnt}), and the corresponding ELISA-equivalent MC concentrations were determined by ELISA (Section~\ref{subsec:MCCnt}). The level-specific reference values were assigned to the eight corresponding imaging replicates. The reference cell concentrations and ELISA-equivalent MC concentrations for the calibration and held-out validation datasets are summarized in Supplementary Table~\ref{tab:sample_reference_values}.

\subsection{Reference Measurements}
\label{subsec:reference_measurements}
\subsubsection{Cell Enumeration}
\label{subsec:cellCnt}
\textit{Microcystis} spp. cell concentration was quantified from the level-specific reference aliquot collected from each homogenized bulk preparation using a hemocytometer under $400\times$ magnification following established cyanobacterial counting procedures~\cite{WHO2021,HotzelCroome1999}. Prior to counting, the reference aliquot was gently homogenized to promote uniform cell distribution, and an appropriate volume was loaded into the hemocytometer chamber and examined using an optical microscope.

Cell counts were performed on fresh, unfixed samples without Lugol's iodine or other chemical fixatives or stains to minimize potential preservation induced changes in cell morphology and volume~\cite{Hawkins2005,Bartram2000}. Cells were counted within predefined grid areas using standard boundary inclusion and exclusion rules~\cite{HotzelCroome1999}. The chamber was independently loaded three times from the homogenized reference aliquot, and the corresponding counts were used to assess within sample counting variability. The mean cell count across the three chamber loads/grid areas was converted to cells~mL$^{-1}$ based on the known volume of the counted chamber area and the applicable dilution factor~\cite{BiologyWG2024}. The resulting cell concentration was assigned as the level-specific reference value for the eight corresponding imaging replicates.

\subsubsection{MC Quantification}
\label{subsec:MCCnt}
Level-specific MC reference aliquots were stored at $-20~^\circ$C and transported in insulated cooled containers with minimal light exposure prior to analysis. MC reference concentrations were determined by the National Water Supply and Drainage Board, Vavuniya, Sri Lanka, using the Beacon MC Plate Kit (Beacon Analytical Systems, Inc., Saco, ME, USA; Cat. No. 20-0068), a competitive ELISA for MCs in water, following the manufacturer's analytical procedure~\cite{BeaconMicrocystinKit}.

The submitted samples were analyzed as whole sample aliquots. Because the assay is calibrated against MC-LR and the kit antibody exhibits cross reactivity with multiple MC congeners, the reported reference values were interpreted in this study as ELISA-equivalent total MC concentrations based on an MC-LR calibration standard~\cite{BeaconMicrocystinKit}.

\subsection{Multispectral Imaging System}
\label{sub:MSI_Dev}
\begin{figure}
    \centering
    \includegraphics[width=0.5\linewidth]{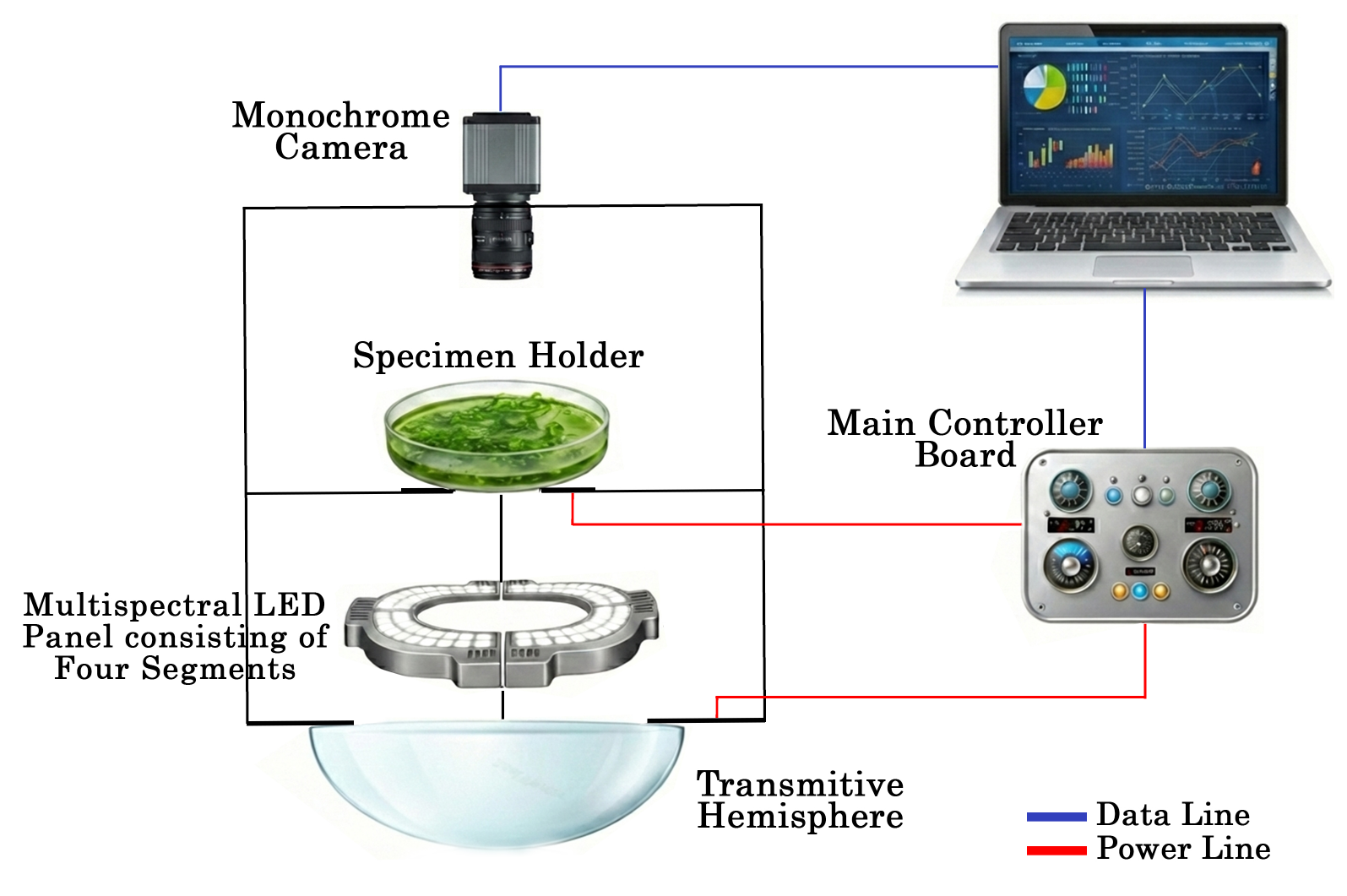}
    \caption{Schematic of the transmission mode multispectral imaging system used for controlled illumination, image acquisition, and synchronized wavelength selection.}
    \label{fig:msi_device}
\end{figure}

A custom in-house transmission mode MSI system, shown in Fig.~\ref{fig:msi_device}, was used to acquire wavelength specific optical measurements from the prepared samples~\cite{Udayanga2024}. The system acquires images at twelve selected spectral bands spanning the near-ultraviolet (NUV) to NIR regions, thereby providing a compact multispectral representation without requiring densely sampled hyperspectral acquisition. Narrowband LEDs with peak wavelengths of 365, 405, 473, 530, 621, 660, 735, 770, 830, 850, 890, and 940~nm were used as the illumination sources. Multispectral image acquisition was performed by sequentially activating each LED. Image capture was carried out using a monochrome industrial machine vision camera (FLIR BFS-U3-13Y3M; 1.3~MP; $1280\times1024$; 10-bit ADC) equipped with an onsemi PYTHON1300 image sensor and a C-mount lens interface. The spectral sensitivity range of the camera extends approximately from 350 to 1080~nm, covering the selected illumination wavelengths.

The camera, illumination panel, and sample holder were enclosed within a custom built dark chamber to minimize the effects of ambient light and external reflections, thereby ensuring repeatable measurements. The interior surfaces were coated with a matte black nonreflective material to suppress stray light and improve imaging consistency across spectral bands. System control and synchronization were achieved using an Arduino Due microcontroller, a controller board, and a host computer. 

\subsection{Image Acquisition}
\label{sub:image_aq}

Prior to multispectral acquisition, the imaging system was configured to provide stable and repeatable image capture conditions. The camera lens system was adjusted by refining the focus and setting the aperture and exposure to maintain adequate signal levels across the wavelength range without sensor saturation. After this initial configuration, the imaging geometry and camera settings were kept fixed for all samples.

Samples were placed at a fixed capture position directly beneath the camera using a custom sample container with opaque sidewalls to suppress stray light interference and a transparent glass base to provide a stable optical interface during imaging. For each sample, the twelve narrowband LEDs were activated sequentially, and one monochrome image was captured at each wavelength, producing a twelve band multispectral image stack. One additional dark frame was acquired for the same sample with all illumination sources switched off and was used as the common dark current reference for the twelve wavelength images.

\subsection{Analysis of Multispectral Images}

\begin{figure}
    \centering
    \includegraphics[width=0.75\linewidth]{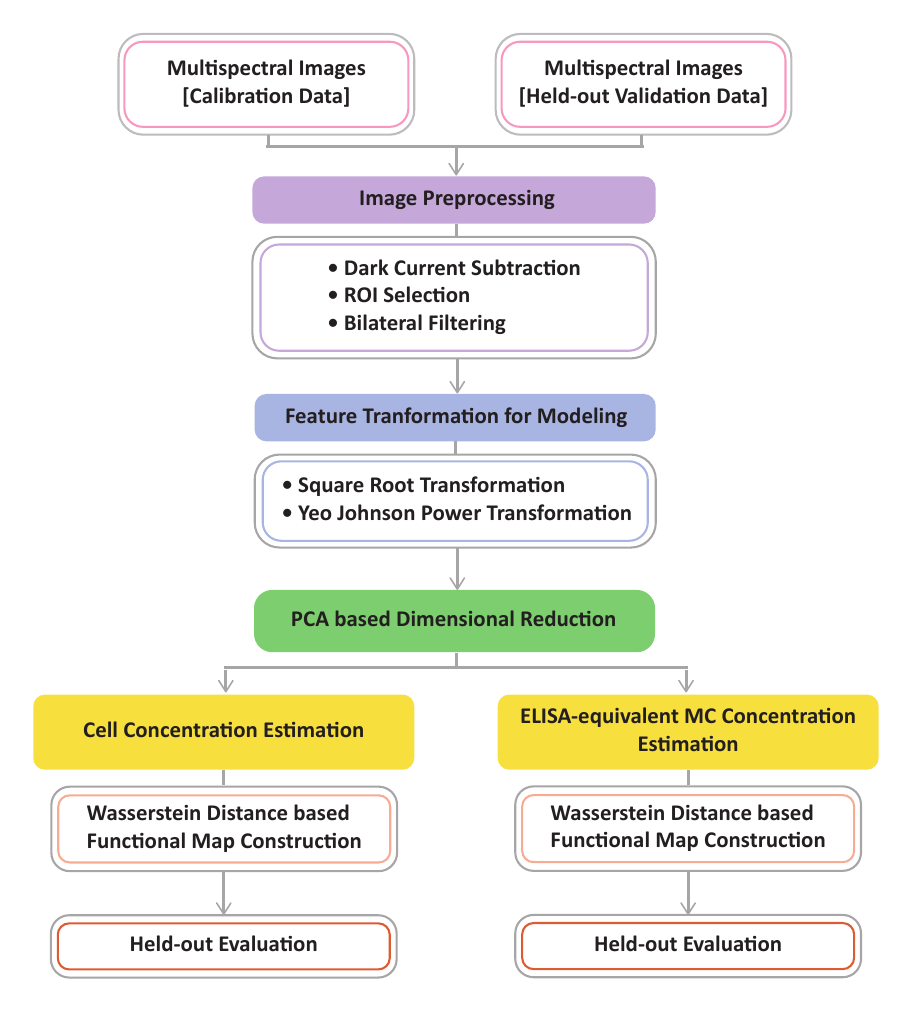}
    \caption{Overview of the multispectral image processing workflow used for \textit{Microcystis} spp. cell concentration and ELISA-equivalent MC concentration estimation. The pipeline includes preprocessing, feature matrix construction, feature transformation, PCA based dimensionality reduction, Wasserstein distance descriptor construction, regression modeling, and evaluation on held-out intermediate dilution levels.}
    \label{fig:flowchart}
\end{figure}

Following multispectral image acquisition, the images underwent several signal-processing stages to estimate \textit{Microcystis} spp. cell concentration and ELISA-equivalent MC concentration, as illustrated in Fig.~\ref{fig:flowchart}. The analysis first involved image preprocessing, including dark current subtraction, region of interest (ROI) selection, and bilateral filtering, to reduce acquisition related noise and standardize the spectral images prior to feature extraction. The preprocessed multispectral images were then converted into structured data matrices. To improve the suitability of the extracted features for subsequent modeling, square root and Yeo-Johnson transformations were applied, followed by PCA based dimensionality reduction to obtain a compact representation of the multispectral information. Finally, functional relationships were established between the reduced spectral representation and the two target variables: \textit{Microcystis} spp. cell concentration and ELISA-equivalent MC concentration.

\subsubsection{Image Preprocessing}
\label{subsub:ImagePreprocessing}

Preprocessing constitutes an important preliminary stage in multispectral image analysis, as it improves data quality and facilitates the extraction of meaningful information from the acquired images. It also helps minimize the effects of noise introduced during image acquisition. Imaging sensors are subject to random noise sources, including dark current noise and fixed pattern noise, even in the absence of illumination~\cite{janesick2007photon}. To mitigate sensor related background contributions, the dark frame $D$ acquired for each sample with all LEDs switched off was used as the common dark-current reference for its twelve illuminated wavelength images. For an image acquired at wavelength $\lambda$, the correction was performed as

\begin{equation}
P[\lambda] = \max(S[\lambda]-D,0),
\end{equation}
where $S[\lambda]$ denotes the image acquired under illumination at wavelength $\lambda$, $D$ denotes the common all LED off dark frame, and $P[\lambda]$ is the corresponding dark corrected transmitted intensity image. The maximum operation was applied element wise to suppress small negative residuals arising from sensor read noise and quantization at low signal levels. No wavelength specific blank or white reference normalization was applied. Therefore, the corrected image values were treated as relative transmitted intensities acquired under fixed imaging conditions rather than absolute radiometric transmittance.

Thereafter, each spectral band image was cropped to a fixed $200\times200$ pixel ROI centered within the sample region. The same spatial ROI was used across all twelve wavelength images of a given sample to minimize edge related effects and maintain spatial correspondence between spectral bands.

Following cropping, to suppress residual random noise while preserving meaningful local intensity variations and textural characteristics of the specimen, a bilateral filter was applied~\cite{tomasi1998bilateral} to each dark current corrected image. The bilateral filter has the capability to perform data adaptive denoising by considering both the spatial proximity and intensity similarity of neighboring pixels. This makes it suitable for multispectral image preprocessing, where noise reduction is required without excessively smoothing informative spatial patterns present in the sample. 

The filter parameters were fixed at a spatial standard deviation of \(\sigma_s=75\), range standard deviation of \(\sigma_r=75\), and neighborhood diameter of 9 pixels. The same parameters were applied to all calibration and held-out validation samples and were not adjusted during evaluation of the held-out set.

\subsubsection{Data Matrix Preparation}
\label{sub:DataPrep}

\begin{figure}
    \centering
    \includegraphics[width=0.95\textwidth]{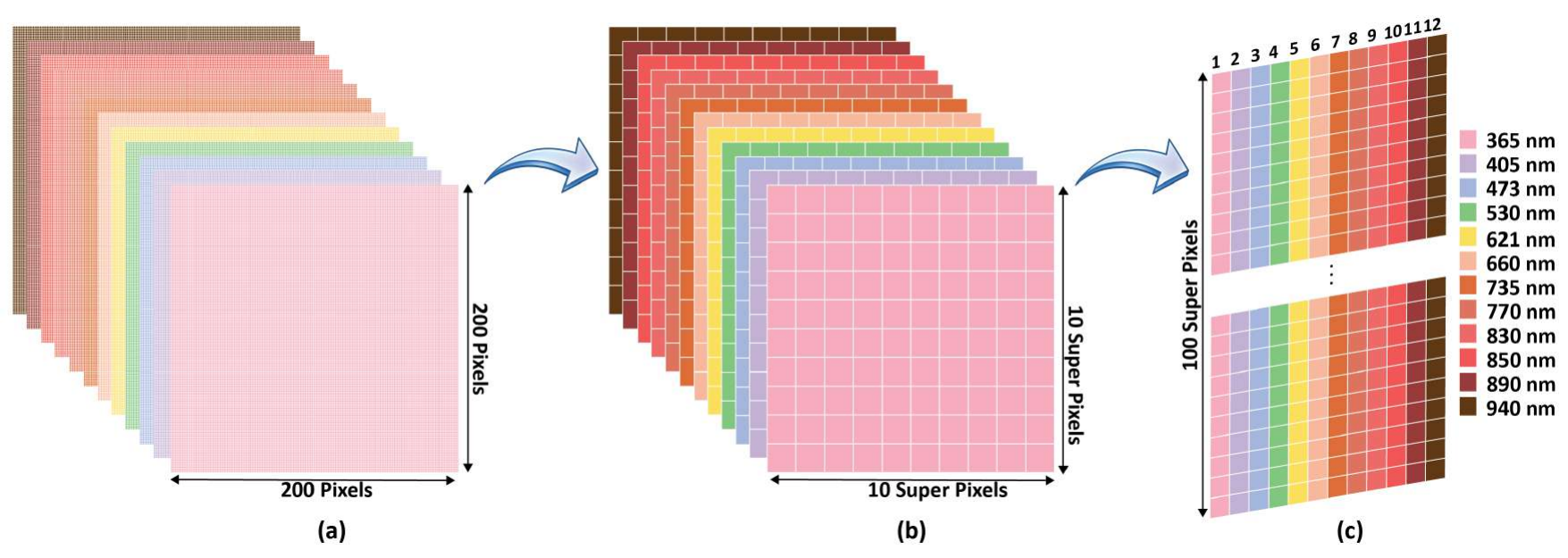}
    \caption{Overview of spectral feature extraction from the preprocessed $200{\times}200$ pixel ROI. The ROI is divided into $20{\times}20$ blocks, and the mean intensity of each block is computed for all twelve wavelength bands to construct the per-sample feature matrix $X_{\mathrm{sample}}\in\mathbb{R}^{100\times12}$.}
    \label{fig:feat_extraction}
\end{figure}

Following preprocessing (Section~\ref{subsub:ImagePreprocessing}), each sample replicate consisted of twelve co-registered monochrome images corresponding to the selected wavelength bands. The fixed $200\times200$ pixel ROI was subdivided into non overlapping $20\times20$ pixel blocks, yielding 100 spatial regions per sample. The mean intensity within each block was computed independently for each wavelength band, as illustrated in Fig.~\ref{fig:feat_extraction}.

For an individual sample replicate, the resulting block averaged spectral values were arranged into the matrix

\begin{equation}
X_{\mathrm{sample}} \in \mathbb{R}^{100\times12},
\end{equation}
where each row represents one spatial block and each column corresponds to one spectral band.

For concentration level $\ell$, eight imaging replicates derived from the same homogenized bulk preparation were available:

\begin{equation}
\left\{
X_{\ell,1},
X_{\ell,2},
\dots,
X_{\ell,8}
\right\},
\qquad
X_{\ell,i}\in\mathbb{R}^{100\times12}.
\end{equation}

The eight sample matrices belonging to the same concentration level were vertically concatenated to form

\begin{equation}
X_{\ell}
=
\begin{bmatrix}
X_{\ell,1}\\
X_{\ell,2}\\
\vdots\\
X_{\ell,8}
\end{bmatrix}
\in\mathbb{R}^{800\times12}.
\end{equation}

Thus, each level wise matrix contained 800 block level observations, corresponding to 100 spatial blocks from each of the eight sample replicates.

The calibration set contained eleven concentration levels. The corresponding level wise matrices were vertically concatenated to construct

\begin{equation}
X_{\mathrm{cal}}
=
\begin{bmatrix}
X_{1}\\
X_{2}\\
\vdots\\
X_{11}
\end{bmatrix}
\in\mathbb{R}^{8800\times12},
\end{equation}
where $8800=11\times8\times100$.

The held-out validation set contained six intermediate concentration levels. Using the same procedure, the held-out matrix was constructed as

\begin{equation}
X_{\mathrm{hold}}
\in\mathbb{R}^{4800\times12},
\end{equation}
where $4800=6\times8\times100$.

The rows of $X_{\mathrm{cal}}$ and $X_{\mathrm{hold}}$ represent block level observations used to characterize the spectral distributions and do not represent independent sample replicates.

Min-max normalization was applied independently to each spectral band. The scaling parameters were estimated using $X_{\mathrm{cal}}$ only and were subsequently applied unchanged to $X_{\mathrm{hold}}$.

\subsubsection{Spectral Signature}
\label{sub:SpecSig}

After data matrix preparation (Section~\ref{sub:DataPrep}), spectral signatures were plotted to provide an initial visualization of the spectral behavior associated with the prepared concentration levels. These signatures were obtained by plotting the normalized mean intensity values against wavelength for the calibration concentration levels (see Fig.~\ref{fig:spectral_sig}).

\subsubsection{Feature Transformation for Modeling}
\label{sub:FeatureTransform}

To further refine the spectral feature representation, additional transformations were applied to the normalized calibration and held-out validation matrices. First, to reduce the dynamic range of the normalized intensities while preserving their relative ordering, an element wise square root transformation was applied to $X_{\mathrm{cal}}$ and $X_{\mathrm{hold}}$:

\begin{equation}
X^{(\mathrm{sqrt})}_{ij}
=
\sqrt{\max(X_{ij},0)},
\qquad
X \in \{X_{\mathrm{cal}},X_{\mathrm{hold}}\}.
\end{equation}
where $(i,j)$ denotes the position of a feature value in the corresponding data matrix. The $\max(\cdot,0)$ operation was included as a safeguard to ensure non negative inputs to the square root operation.

Subsequently, a Yeo-Johnson power transform was applied independently to each wavelength feature to reduce skewness, stabilize variance, and produce a more well behaved feature distribution for the subsequent dimensionality reduction step in Section~\ref{subsec:dimRed}~\cite{yeo2000power}. The feature specific transformation parameters were estimated from the calibration data, $X_{\mathrm{cal}}$ by maximum likelihood estimation and were then fixed and applied unchanged to the corresponding features in the held-out validation data, $X_{\mathrm{hold}}$.

A representative example illustrating the feature distributions after min-max normalization, square root transformation, and Yeo-Johnson power transformation is provided in Supplementary Fig.~\ref{fig:transformation_hist_890}.

\subsubsection{Dimension Reduction}
\label{subsec:dimRed}

Captured multispectral image data are not entirely uncorrelated because of spectral overlap among adjacent wavelength bands. To obtain a more compact spectral feature representation, PCA was employed as an unsupervised dimensionality reduction technique~\cite{jolliffe2016pca}. PCA transforms the original correlated spectral bands into a new set of orthogonal features, referred to as principal components, while preserving most of the variance present in the data.

Following the feature transformation step described in Section~\ref{sub:FeatureTransform}, PCA was fitted using only the transformed calibration matrix. Let $X_{\mathrm{cal}}\in\mathbb{R}^{8800\times12}$ denote the transformed calibration matrix, where the 8800 rows correspond to block level spectral observations obtained from eleven calibration levels, eight sample replicates per level, and 100 spatial blocks per sample. Similarly, let $X_{\mathrm{hold}}\in\mathbb{R}^{4800\times12}$ denote the transformed held-out validation matrix.

The calibration matrix was mean centered using its mean vector, $\boldsymbol{\mu}\in\mathbb{R}^{12}$, and the covariance matrix of the centered data was computed. Eigenvalue decomposition of the covariance matrix was then used to identify the principal directions of spectral variation. The corresponding eigenvectors were ordered by decreasing eigenvalue, and the first $k$ eigenvectors were used to construct the projection matrix
\begin{equation}
W =
[\mathbf{v}_1,\mathbf{v}_2,\dots,\mathbf{v}_k],
\end{equation}
where $\mathbf{v}_{\ell}$ denotes the eigenvector associated with the $\ell$ th largest eigenvalue. The number of retained components, $k$, was selected to preserve at least 99\% of the total variance in the calibration data. This criterion resulted in the selection of the first four principal components, giving $W\in\mathbb{R}^{12\times4}$. Detailed expressions for mean centering, covariance estimation, and eigenvalue decomposition are provided in the Supplementary under Mathematical formulation of PCA.

The reduced representation of the calibration data was obtained as
\begin{equation}
Z_{\mathrm{cal}}
=
\left(
X_{\mathrm{cal}}-\mathbf{1}\boldsymbol{\mu}^{\top}
\right)W,
\end{equation}
where $Z_{\mathrm{cal}}\in\mathbb{R}^{8800\times4}$ and $\mathbf{1}$ denotes a column vector of ones with length equal to the number of rows in the corresponding data matrix.

The calibration derived mean vector $\boldsymbol{\mu}$ and projection matrix $W$ were then applied unchanged to the held-out validation data:
\begin{equation}
Z_{\mathrm{hold}}
=
\left(
X_{\mathrm{hold}}-\mathbf{1}\boldsymbol{\mu}^{\top}
\right)W,
\end{equation}
where $Z_{\mathrm{hold}}\in\mathbb{R}^{4800\times4}$. Thus, both datasets were represented in the same four dimensional PCA space without refitting PCA to the held-out observations.

\begin{figure*}
    \centering
    \includegraphics[width=1\linewidth]{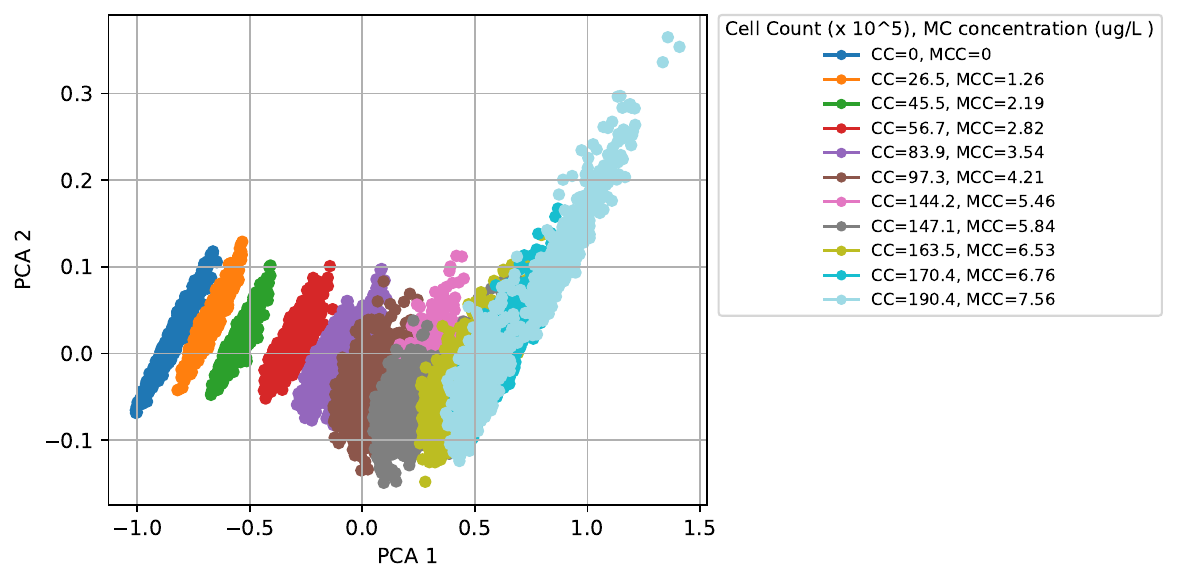}
    \caption{PCA representation of the transformed calibration data projected onto the first two principal components. Each point represents a block level spectral observation and is grouped according to its corresponding concentration level.}
    \label{fig:PCAplot}
\end{figure*}

To visualize the structure of the reduced feature space, the calibration observations projected onto the first two principal components were plotted, as shown in Fig.~\ref{fig:PCAplot}. The block level observations exhibit clustering according to concentration level, with partial overlap among several adjacent levels. This indicates that the PCA transformed features retain concentration dependent spectral variation for subsequent modeling.

\subsubsection{Wasserstein Distance Based Regression Model}
\label{subsec:RegMod}

To establish a functional relationship between each target variable and a scalar spectral descriptor, the Wasserstein distance was employed as a measure of distributional separation~\cite{villani2009optimal,rubner2000emd}. In this study, the 1-Wasserstein distance, $W_1$, was used. Its general mathematical formulation is provided in the Supplementary under Mathematical formulation of the Wasserstein distance.

After PCA, each MSI sample replicate was represented by 100 observations in the four dimensional retained PCA space. One dimensional marginal Wasserstein distances were calculated independently along each retained principal component relative to a single fixed M8 medium reference sample from the 0\% concentration level, rather than directly computing a multivariate Wasserstein distance.

Let $Z_{\mathrm{ref}}^{(k)}$ denote the distribution of the 100 PCA scores of the fixed M8 medium reference sample along the $k$ th principal component, and let $Z_{i}^{(k)}$ denote the corresponding distribution for MSI sample $i$. The component wise Wasserstein distance was calculated as

\begin{equation}
w_{i}^{(k)}
=
W_1
\left(
Z_{\mathrm{ref}}^{(k)},
Z_{i}^{(k)}
\right),
\qquad
k=1,\ldots,4.
\end{equation}

The four component wise Wasserstein distances were then combined using an unweighted sum to obtain a single scalar descriptor for sample $i$:

\begin{equation}
\mathcal{W}_{i}
=
\sum_{k=1}^{4} w_{i}^{(k)}.
\end{equation}

Thus, each MSI sample was represented by one aggregated Wasserstein distance descriptor, $\mathcal{W}_{i}$. 

The same fixed M8 medium reference sample from the 0\% dilution level was used for both the calibration and held-out validation samples.

This procedure produced 88 sample level Wasserstein descriptors for the calibration set, corresponding to eleven concentration levels with eight MSI sample replicates per level, and 48 sample level descriptors for the held-out validation set, corresponding to six intermediate concentration levels with eight MSI sample replicates per level.
Each sample level descriptor was paired with the reference concentration assigned to its corresponding dilution level. Thus, the 88 calibration descriptors were associated with 11 distinct level wise reference concentrations, while the 48 held-out descriptors were associated with six distinct level wise reference concentrations. 

Separate linear regression models were fitted between the aggregated Wasserstein distance descriptor and the two target variables. For \textit{Microcystis} spp. cell concentration,

\begin{equation}
\hat{C}_{\mathrm{cell},i}
=
\beta_{0,\mathrm{cell}}
+
\beta_{1,\mathrm{cell}}
\mathcal{W}_{i},
\end{equation}

whereas for ELISA-equivalent MC concentration,

\begin{equation}
\hat{C}_{\mathrm{MC},i}
=
\beta_{0,\mathrm{MC}}
+
\beta_{1,\mathrm{MC}}
\mathcal{W}_{i}.
\end{equation}

The regression coefficients were estimated using only the calibration data. The fitted models were subsequently evaluated on the held-out validation samples.

\subsection{Uncertainty and Error Analysis}
\label{subsec:uncertainty}

To characterize prediction uncertainty and sample level error for the two linear calibration models defined in Section~\ref{subsec:RegMod}, uncertainty and error analyses were performed separately for \textit{Microcystis} spp. cell concentration and ELISA-equivalent MC concentration. The regression models and associated uncertainty bands were estimated using only the calibration data. The fitted models were then applied unchanged to the held-out intermediate dilution levels. For each target, 95\% confidence intervals were used to describe uncertainty in the mean fitted calibration response, while 95\% prediction intervals were used to characterize the expected dispersion of individual sample level predictions around the fitted relationship.

Prediction error for sample $i$ was calculated as

\begin{equation}
e_i
=
\hat{C}_i-C_i,
\end{equation}

where $\hat{C}_i$ is the MSI predicted concentration and $C_i$ is the reference concentration assigned to the corresponding sample. Positive errors indicate over prediction, whereas negative errors indicate under prediction. Bias was calculated as the mean prediction error, and residual standard deviation was calculated as the sample standard deviation of the prediction errors.

Prediction performance was summarized using the $R^2$, RMSE, MAE, bias, residual standard deviation, and 95\% prediction interval coverage. These quantities were evaluated at the MSI sample level, using one aggregated Wasserstein distance descriptor per imaging replicate. Accordingly, the calibration statistics were based on 88 sample level descriptors associated with 11 distinct concentration levels, whereas the held-out statistics were based on 48 sample level descriptors associated with six distinct intermediate concentration levels.

\section*{Code availability}

The code used for multispectral image preprocessing, feature transformation, principal component analysis, Wasserstein distance descriptor construction, regression modelling, and performance evaluation is publicly available at \url{https://github.com/SenithJayakody/microcystis-microcystin-msi}.

\bibliography{sn-bibliography}
\section*{Acknowledgements}

The authors acknowledge the Department of Botany, Faculty of Science, University of Peradeniya, Sri Lanka, for providing the freshwater bloom sample and assistance with culture preparation and laboratory procedures. The authors also thank the National Water Supply and Drainage Board, Vavuniya, Sri Lanka, for supporting the microcystin analysis.

\section*{Funding}

This research was supported by the National Research Council of Sri Lanka under the Investigator-Driven Research Grant (Grant No. 25-015).

\clearpage

\pagestyle{empty}
\thispagestyle{empty}

\begin{center}
{\Large\bfseries Supplementary Information\par}
\vspace{1em}
{\large\bfseries
Transmission-Mode Multispectral Imaging for
Estimation of Microcystis spp. Cell Concentration
and ELISA-Equivalent Microcystin Concentration
in Water\par}
\vspace{0.7em}
J.A.S.T. Jayakody, G. Thilakarathne,
G.A.S.L. Ranasinghe, M.C.L. De Silva, K.A.D.S.A. Perera, G.M.R.I. Godaliyadda, H.M.V.R. Herath, M.P.B. Ekanayake, A.D.S.N.P. Athukorala, and H.A.D.R.P. Sirimanna.
\end{center}

\vspace{1em}

\begin{appendices}

\setcounter{equation}{0}
\renewcommand{\theequation}{S\arabic{equation}}
\renewcommand{\theHequation}{supp.\arabic{equation}}

\setcounter{figure}{0}
\renewcommand{\thefigure}{S\arabic{figure}}
\renewcommand{\theHfigure}{supp.\arabic{figure}}

\setcounter{table}{0}
\renewcommand{\thetable}{S\arabic{table}}
\renewcommand{\theHtable}{supp.\arabic{table}}

\subsection*{Growth Characteristics of \textit{Microcystis} spp. Under Laboratory Cultivation}
\label{app:growth_profile}

\begin{figure}[H]
    \centering
    \includegraphics[width=0.9\linewidth]{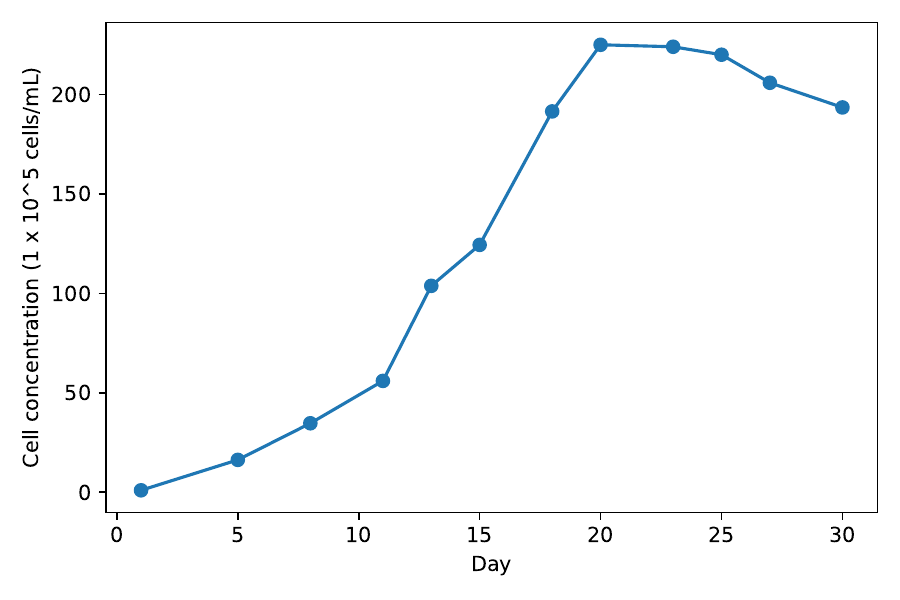}
    \caption{Growth profile of the cultivated \textit{Microcystis} spp. over a 30 day period based on hemocytometer counts obtained at selected time points. The curve was used to identify the stationary growth region for preparation of the controlled concentration series.}
    \label{fig:growth_curve}
\end{figure}

The cultivation dynamics of \textit{Microcystis} spp. were evaluated over a 30 day period under the experimental conditions detailed in Section~\ref{subsec:culture_isolation}.
The observed variations in cell concentration provided an indication of the successive stages of culture development, encompassing the lag, exponential, stationary, and initial decline phases. The temporal growth pattern, illustrated in Fig.~\ref{fig:growth_curve}, served as
the basis for identifying the stationary phase cultivation period used to obtain the culture material for subsequent preparation of the controlled concentration series.

\subsection*{Reference Cell Concentrations and ELISA-Equivalent MC Concentrations}
\label{app:reference_values}

Reference cell concentrations were determined using the hemocytometer based enumeration procedure described in Section~\ref{subsec:cellCnt}, while ELISA-equivalent microcystin (MC) concentrations were obtained as described in Section~\ref{subsec:MCCnt}. The level-specific reference values were assigned to the eight corresponding multispectral imaging replicates at each concentration level. The reference cell concentrations and ELISA-equivalent MC concentrations for the calibration and held-out validation datasets are presented in Table~\ref{tab:sample_reference_values}.

\begin{table}[htbp]
\caption{Prepared dilution levels and corresponding reference cell concentrations and ELISA-equivalent MC concentrations for the calibration and held-out validation datasets.}
\label{tab:sample_reference_values}

\centering
\small
\setlength{\tabcolsep}{5pt}
\renewcommand{\arraystretch}{1.15}

\begin{tabular}{@{}l c r r@{}}
\hline
\textbf{Dataset} &
\shortstack{\textbf{Dilution level}\\\textbf{(\%, v/v)}} &
\shortstack{\textbf{Cell concentration}\\\textbf{($\times 10^{5}$ cells mL$^{-1}$)}} &
\shortstack{\textbf{ELISA-equivalent MC}\\\textbf{($\mu$g L$^{-1}$)}} \\
\hline

Calibration & 0   & 0.00   & 0.00 \\
Calibration & 10  & 26.50  & 1.26 \\
Calibration & 20  & 45.50  & 2.19 \\
Calibration & 30  & 56.70  & 2.82 \\
Calibration & 40  & 83.90  & 3.54 \\
Calibration & 50  & 97.30  & 4.21 \\
Calibration & 60  & 144.20 & 5.46 \\
Calibration & 70  & 147.10 & 5.84 \\
Calibration & 80  & 163.50 & 6.53 \\
Calibration & 90  & 170.40 & 6.76 \\
Calibration & 100 & 190.40 & 7.56 \\

\hline

Held-out validation & 5  & 19.40  & 0.62 \\
Held-out validation & 15 & 44.40  & 1.90 \\
Held-out validation & 35 & 68.40  & 3.11 \\
Held-out validation & 55 & 131.10 & 4.94 \\
Held-out validation & 75 & 158.10 & 6.29 \\
Held-out validation & 95 & 181.70 & 6.98 \\

\hline
\end{tabular}

\vspace{2mm}

\begin{minipage}{\linewidth}
\footnotesize
\textit{Note:} Cell concentrations and ELISA-equivalent MC concentrations are level-specific reference values determined from aliquots collected from each homogenized bulk preparation before subdivision into eight imaging replicates. The same level-specific reference values were therefore assigned to the eight corresponding MSI imaging replicates at each dilution level.
\end{minipage}

\end{table}

\subsection*{Illustration of Spectral Feature Transformations}
\label{app:feature_transformations}

\begin{figure}[H]
    \centering
    \includegraphics[width=0.75\textwidth]{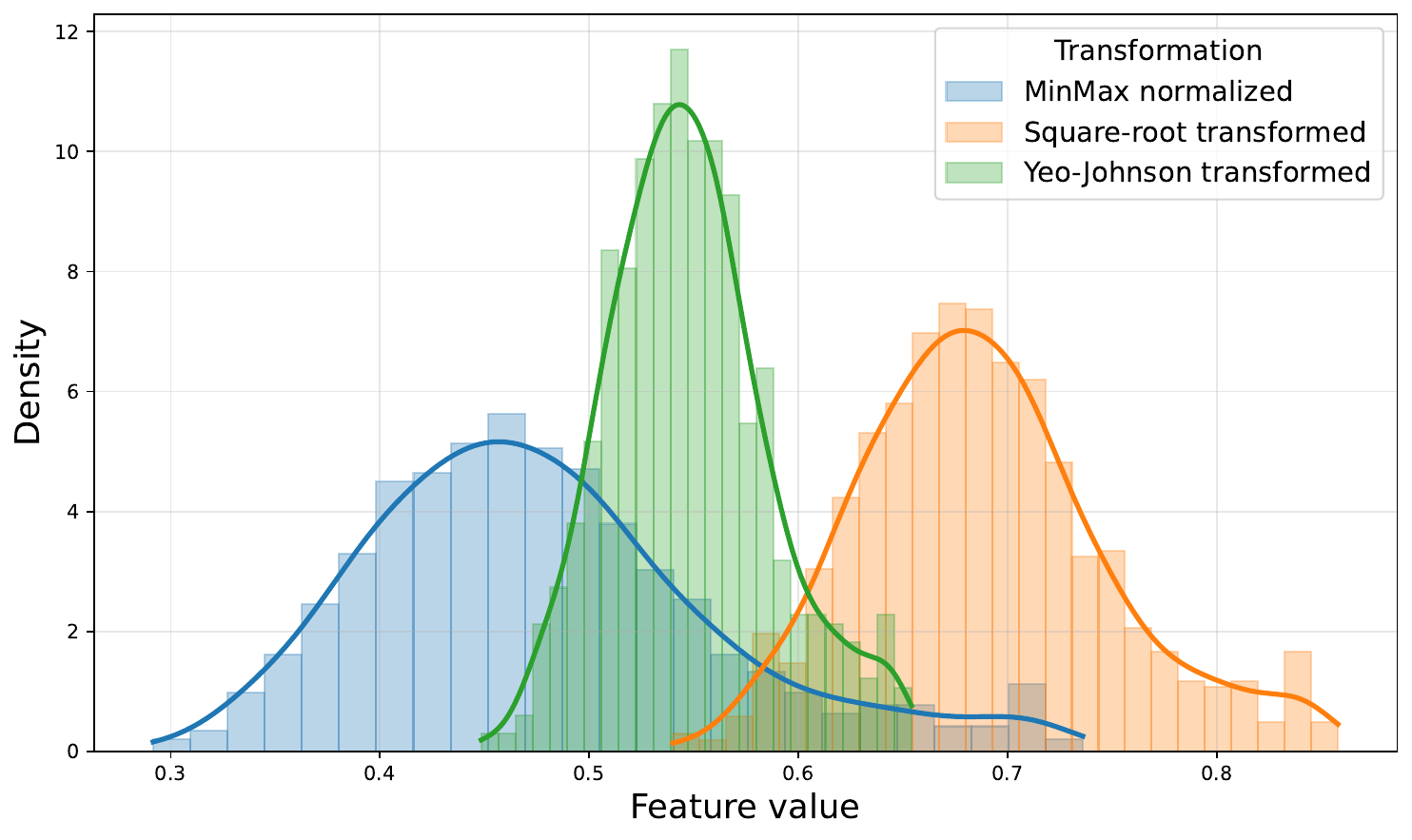}
    \caption{Representative histogram of the 890~nm wavelength feature at the concentration level of $56.7 \times 10^5$ cells mL$^{-1}$, corresponding to an ELISA-equivalent MC concentration of 2.82~$\mu$g L$^{-1}$. The figure shows the feature distribution after min-max normalization, square-root transformation, and Yeo-Johnson power transformation.}
    \label{fig:transformation_hist_890}
\end{figure}

To illustrate the effect of the sequential feature transformations, Fig.~\ref{fig:transformation_hist_890} shows a representative example for the 890~nm wavelength feature at the concentration level of $56.7 \times 10^5$ cells mL$^{-1}$, corresponding to an ELISA-equivalent MC concentration of 2.82~$\mu$g L$^{-1}$. The distributions after min-max normalization, square root transformation, and Yeo-Johnson power transformation are shown to demonstrate how the preprocessing steps modified the feature distribution.

\subsection*{Mathematical formulation of PCA}
\label{app:PCA}

This section provides the mathematical details of the Principal Component Analysis (PCA) used for dimensionality reduction in Section~\ref{subsec:dimRed}.

Let $X_{\mathrm{cal}}\in\mathbb{R}^{N_{\mathrm{cal}}\times12}$ denote the transformed calibration matrix, where $N_{\mathrm{cal}}=8800$. Each row contains the twelve transformed spectral features of one spatial block. Let $\mathbf{x}_i\in\mathbb{R}^{12}$ denote the column vector corresponding to the $i$ th row of $X_{\mathrm{cal}}$.

The calibration mean vector was calculated as
\begin{equation}
\boldsymbol{\mu}
=
\frac{1}{N_{\mathrm{cal}}}
\sum_{i=1}^{N_{\mathrm{cal}}}
\mathbf{x}_i.
\label{eq:supp_pca_mean}
\end{equation}

The calibration matrix was then mean centered:
\begin{equation}
\widetilde{X}_{\mathrm{cal}}
=
X_{\mathrm{cal}}
-
\mathbf{1}\boldsymbol{\mu}^{\top},
\label{eq:supp_pca_centering}
\end{equation}
where $\mathbf{1}\in\mathbb{R}^{N_{\mathrm{cal}}}$ is a column vector of ones.

The sample covariance matrix, $C\in\mathbb{R}^{12\times12}$, was computed from the centered calibration data as
\begin{equation}
C
=
\frac{1}{N_{\mathrm{cal}}-1}
\widetilde{X}_{\mathrm{cal}}^{\top}
\widetilde{X}_{\mathrm{cal}}.
\label{eq:supp_pca_covariance}
\end{equation}

The principal directions were obtained through eigenvalue decomposition:
\begin{equation}
C\mathbf{v}_{\ell}
=
\lambda_{\ell}\mathbf{v}_{\ell},
\qquad
\ell=1,\ldots,12,
\label{eq:supp_pca_eigendecomposition}
\end{equation}
where $\lambda_{\ell}$ and $\mathbf{v}_{\ell}$ denote the corresponding eigenvalue and unit length eigenvector. The eigenvectors were chosen to be mutually orthogonal and ordered according to decreasing eigenvalue,
$\lambda_1\geq\lambda_2\geq\cdots\geq\lambda_{12}\geq0$.
Each eigenvalue represents the variance along its corresponding principal component.

The number of retained principal components was selected as the smallest number required to preserve at least 99\% of the total calibration variance:
\begin{equation}
k
=
\min
\left\{
m\in\{1,\ldots,12\}:
\frac{\sum_{\ell=1}^{m}\lambda_{\ell}}
     {\sum_{\ell=1}^{12}\lambda_{\ell}}
\geq 0.99
\right\}.
\label{eq:supp_pca_variance}
\end{equation}

This criterion yielded $k=4$. The corresponding eigenvectors formed the projection matrix
$W=[\mathbf{v}_1,\ldots,\mathbf{v}_4]\in\mathbb{R}^{12\times4}$.
The calibration data were projected using this matrix after mean centering.

For the held-out validation matrix,
$X_{\mathrm{hold}}\in\mathbb{R}^{4800\times12}$,
centering and projection were performed using the same calibration mean vector $\boldsymbol{\mu}$ and projection matrix $W$. Neither quantity was re-estimated from the held-out data. The resulting reduced matrices had dimensions
$Z_{\mathrm{cal}}\in\mathbb{R}^{8800\times4}$ and
$Z_{\mathrm{hold}}\in\mathbb{R}^{4800\times4}$.

\subsection*{Mathematical formulation of the Wasserstein distance}
\label{app:wasserstein}

The Wasserstein distance quantifies the dissimilarity between two probability distributions in terms of the minimum cost required to transport probability mass from one distribution to another, thereby capturing both shifts in location and differences in spread of the underlying feature distributions.

Let $\mu$ and $\nu$ be two probability measures on $\mathbb{R}^{d}$, and let $\Pi(\mu,\nu)$ denote the set of all couplings (transport plans) whose marginals are $\mu$ and $\nu$. The $p$-Wasserstein distance is defined as

\begin{equation}
W_p(\mu,\nu)
=
\left(
\inf_{\gamma\in\Pi(\mu,\nu)}
\int_{\mathbb{R}^{d}\times\mathbb{R}^{d}}
\|x-y\|^p\,d\gamma(x,y)
\right)^{1/p}.
\end{equation}

In this study, $p=1$ was used, yielding the 1-Wasserstein distance, $W_1$, which is commonly referred to as the Earth Mover's Distance.

\end{appendices}

\end{document}